\documentclass[12pt]{article}

\usepackage{color}
\usepackage{natbib}
\usepackage{graphicx}

\begin{document}

\title{A recent tipping point in the Arctic sea-ice cover:
abrupt and persistent increase in the seasonal cycle since 2007}

\date{}

\maketitle

\vspace{-20mm}

\begin{center}
{Valerie N. Livina$^{1,2}$ and Timothy M. Lenton$^3$}\\[10pt]
{\it $^1$National Physical Laboratory, Teddington, Middlesex TW11 0LW, UK}\\
{\it $^2$School of Environmental Sciences, University of East Anglia, Norwich NR4 7TJ, UK}\\
{\it $^3$College of Life and Environmental Sciences, University of Exeter,
Hatherly Laboratories, Exeter EX4 4PS, UK}
\end{center}

\begin{abstract}
There is ongoing debate over whether Arctic sea-ice has already passed a `tipping point',
or whether it will do so in the future.
Several recent studies argue that the loss of summer sea ice does not involve an irreversible
bifurcation, because it is highly reversible in models.
However, a broader definition of a `tipping point' also includes other abrupt, non-linear changes
that are neither bifurcations nor necessarily irreversible.
Examination of satellite data for Arctic sea-ice area reveals an abrupt increase in the
amplitude of seasonal variability in 2007 that has persisted since then.
We identified this abrupt transition using recently developed methods
that can detect multi-modality in time-series data and sometimes forewarn of bifurcations.
When removing the mean seasonal cycle (up to 2008) from the satellite data,
the residual sea-ice fluctuations switch from uni-modal to multi-modal behaviour around 2007.
We originally interpreted this as a bifurcation in which a new lower ice cover attractor appears
in deseasonalised fluctuations and is sampled in every summer-autumn from 2007 onwards.
However, this interpretation is clearly sensitive to how the seasonal cycle is removed from the raw data,
and to the presence of continental land masses restricting winter-spring ice fluctuations.
Furthermore, there was no robust early warning signal of critical slowing down
prior to the hypothesized bifurcation.
Early warning indicators do however show destabilization of the summer-autumn
sea-ice cover since 2007.
Thus, the bifurcation hypothesis lacks consistent support,
but there was an abrupt and persistent increase
in the amplitude of the seasonal cycle of Arctic sea-ice cover in 2007,
which we describe as a (non-bifurcation) `tipping point'.
Our statistical methods detect this `tipping point' and its time of onset.
We discuss potential geophysical mechanisms behind it,
which should be the subject of further work with process-based models.
\end{abstract}

\medskip

\begin{figure}[h!]
\centerline{\includegraphics[width=.9\textwidth]{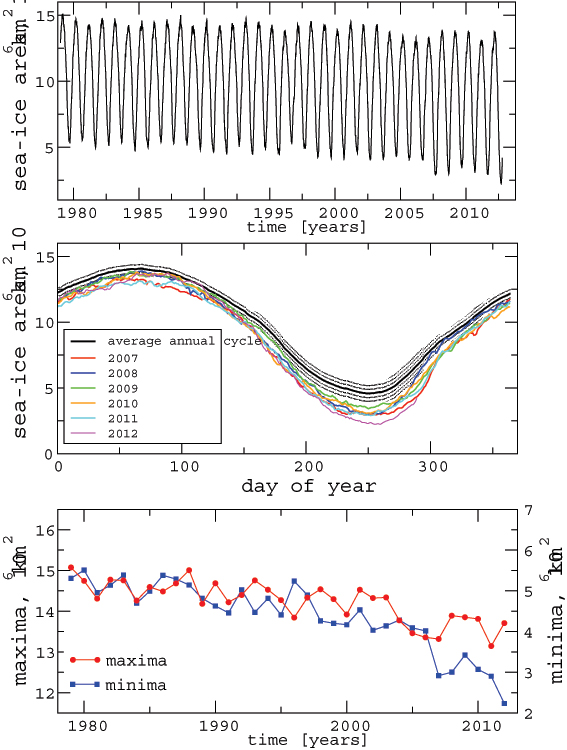}}
\caption{Arctic sea-ice area from satellite data. (a) Arctic sea-ice area, 1979-2012.
(b) The mean annual cycle of the area data over 1979-2008 inclusive (solid line,
shaded area denotes two error bars), together with the last five anomalous years.
(c) Annual maxima (left axis) and minima (right axis) showing an abrupt increase in
amplitude of the seasonal cycle in 2007.}
\label{f1}
\end{figure}

\noindent
{\bf Introduction.}
Arctic sea-ice has experienced striking reductions in areal coverage~\citep{b1,b2},
especially in recent summers, with 2007-2012 having the six lowest ice
cover minima in the satellite record (Figure~\ref{f1}).
Observations have fallen below IPCC model projections~\citep{b1}, despite
the models having been in agreement with the observations in the 1970s. 
The latest models are more consistent with satellite observations (1979-present),
but still fail to capture the full extent of the observed downward trend~\citep{stroeve12}. 
Summer ice cover is forecast to disappear later this century~\citep{b3},
but the nature of the underlying transition is debated~\citep{b4,b5,b6,b7,b8,b9}.

Arctic sea-ice has been identified as a potential tipping element in the
Earth's climate system~\citep{b4},
and at least one study suggests it has already passed a `tipping point'~\citep{b5}.
In the future, some models forecast abrupt ice loss events~\citep{b6},
on the way to a seasonally
ice-free Arctic.
These may qualify as passing tipping points following
the broad definition given in~\citep{b4}
of a point at which a small change in forcing leads to a qualitative change in
the future state of a system.
The definition includes both reversible and irreversible transitions,
bifurcations and some non-bifurcation phenomena.

However, most recent papers on the Arctic sea-ice opt for a narrower definition
of a tipping point, addressing whether summer sea-ice loss will involve an irreversible
(e.g. saddle-node/fold) bifurcation.
They find instead that in models the loss of summer sea-ice
cover is highly reversible~\citep{b6,b7,b8,b9}.
Abrupt ice loss events are then attributed to the loss of year-round
sea-ice in the Arctic making the
remaining ice more vulnerable to summer melt, and prone to larger
fluctuations in area coverage~\citep{b22}.
An exception is a recent model~\citep{b23}
showing that positive feedbacks involving clouds can create multiple stable states
for seasonal ice cover and bifurcations between them.
Furthermore, models of past abrupt climate changes in the Arctic have shown
multiple stable states for sea-ice cover in the Barents and
Kara Seas region and abrupt switches between them~\citep{b20,b21}.
This suggests that sub-Arctic scale `tipping points' in sea-ice cover are conceivable.

On viewing the satellite-derived daily record of sea-ice area
from 1979 to present (Figure~\ref{f1}a),
it is clear that the last five years have been characterized by an increase
in the amplitude of seasonal sea-ice variation (Figure~\ref{f1}b).
The annual ice cover minimum dropped the order of
$\sim$10$^{6}$ km$^{2}$ more than the annual maximum
in 2007 and the difference has been maintained since then (Figure~\ref{f1}c).
This already suggests an abrupt and persistent change in sea-ice dynamics.
It led us to hypothesize that the sea-ice may have passed a bifurcation-type tipping point,
in which a new attractor for lower summer-autumn sea-ice
cover became stable and began to be sampled in summer 2007,
and in every summer since, with seasonal
switches to/from the pre-existing attractor --
see Figure 10 of~\citep{cryo}.

We arrived at this hypothesis by applying recently developed
methods of time-series analysis that can detect
changes in the modality of data~\citep{b10,b11,cimatoribus}
and in some cases forewarn of bifurcations~\citep{b15,b12,b13,b14}.
Our analysis concentrates on the satellite-derived daily record of
sea-ice area from 1979 to 2011 (Figure~\ref{f1}a),
and is repeated on the shorter record of sea-ice extent 1979-2009~\citep{b24}
(Figure~\ref{s1}) in the Appendix.
For much of our analysis, a mean seasonal cycle
(Figure~\ref{f1}b) averaged over the period
1979-2008 was removed from the data, because there is a very
strong seasonally-forced variation in sea-ice area.
The averaged seasonal cycle of the Arctic sea-ice 1979-2008 (Figure~\ref{f1}b)
is very close to a sine wave (no asymmetry over seasons),
and we were interested in studying the behavior of fluctuations
from this typical state of seasonal variation.

However, our interpretation in terms of a changing number of
sea-ice attractors~\citep{cryo}, can be profoundly altered
by changing the interval that is considered the baseline state for the sea-ice~\citep{ditlev}.
The central problem is that any residual seasonal cycle remaining in the data appears bi-modal,
and if there has been a jump in the amplitude of the seasonal cycle around 2007,
it is not possible to remove one average seasonal cycle from the whole record and get rid of all
the residual seasonality~\citep{ditlev}.
We examine this further here with a stochastic model
of growing amplitude in the seasonal cycle,
by comparing analysis of the artificial data from this model with that of the observed data.

The paper is organized as follows:
Section 2 details the data pre-processing and the methods applied.
Section 3 presents the results and discusses them, including some tentative geophysical interpretation.
Section 4 concludes.

\section{Methodology}

\subsection{Data and pre-processing}
Sea-ice area takes into account the fraction of a grid cell
that is covered by sea ice, and can be biased low, especially in summer
when melt ponds are present. Sea-ice extent assumes that any grid point with more
than a certain per cent (for instance, 15\%)
sea ice concentration is totally covered.

Sea ice area data were obtained from `The Cryosphere Today' project of the University of Illinois.
This dataset\footnote{%
http://arctic.atmos.uiuc.edu/cryosphere/timeseries.anom.1979-2008} uses SSM/I and
SMMR series satellite products and spans 1979 to present at daily resolution.
The most recent data in this series is derived from
the Near-Real-Time DMSP SSM/I-SSMIS Daily Polar Gridded Sea Ice Concentrations of
the National Snow \& Ice Data Centre (NSIDC), see~\citep{maslanik}.

The sea-ice extent time series was derived by~\citep{b24}
(data available from ftp://ftp.agu.org/apend/gl/2010gl043741)
on the basis of sea ice concentration
using the NASA Team algorithm from Nimbus-7 SMMR (1978-1987), DMSP SSM/I
(1987-2009), and DMSP SSMIS (2008-present) satellite
passive microwave radiances on a 25km x 25km
polar stereographic grid~\citep{cavalieri,meier,maslanik}.
During periods of instrumental transitions, the overlapping
datasets were averaged. Extent
was calculated by summing the areas of all grid boxes with at
least 15\% ice concentration. Details of the spatial
data interpolation are given by~\citep{b24}.
The time series
spans 1979-2009, and where it has 2-day resolution
(when SMMR operated every other day in three months during the record, in 10/1978,
12/1987 and 1/1988), we interpolate to daily resolution to obtain a homogeneous time-series.

For both datasets -- area and extent --
the mean seasonal cycle over the first 30 years of data (1979-2008) was removed,
as on `The Cryosphere Today' website (and widely reproduced elsewhere).
We also examined the effect of constructing
and removing a different averaging interval (1979-2011),
which produces a very similar residual series,
just vertically shifted along the y-axis,
i.e. the dynamics of the residual fluctuations remained the same.
Hence this gives similar results and we do not show it here.

We also analyzed a derived index of `equivalent sea-ice extent'~\citep{b24}
(available from ftp://ftp.agu.org/apend/gl/2010gl043741),
which is based on the latitude of the sea-ice edge
where it is free to migrate, converted to an area, assuming there were no continents present.

\subsection{Potential analysis}
To detect any multi-modality
in the sea-ice residual data, we use a recently-developed~\citep{b10,b11} and
blind-tested~\citep{b25} method of `potential analysis'. This assumes that
a system is experiencing sufficient short-term stochastic variability (noise)
that it is sampling all of its available states or attractors (given a sufficiently
long time window). Then we take advantage of the fact that the stationary
probability distribution of the resulting data is directly related to the
shape of the underlying potential, which describes the number of underlying
attractors and their stability~\citep{b11}. Thus, with a sufficiently long time window of data
one can deduce the number of attractors and their relative stability or instability.

The time series are modelled by the stochastic differential equation:
\begin{equation}
\dot z(t)=-U'(z)+\sigma\eta,									
\label{eq1}
\end{equation}
where $U$ is a polynomial potential of even order,
$\eta$ is a Gaussian white noise process of unit variance.
Equation (\ref{eq1}) has a corresponding Fokker-Planck equation describing the
probability density function, and crucially this has a stationary solution that
depends only on the underlying potential function and the noise level, $\sigma$;
\begin{equation}
p(z) \sim \exp\frac{-2U(z)}{\sigma^2}.
\label{eq2}
\end{equation}

This allows the underlying potential to be reconstructed from a kernel probability
distribution of time-series data (and an estimate of the noise level) as:
\begin{equation}
U(z) = - \frac{\sigma^2}2 \log p_d(z),									
\label{eq3}
\end{equation}
where $p_d$ is the empirical probability density of the data.

\begin{figure}[h!]
\centerline{\includegraphics[width=.9\textwidth]{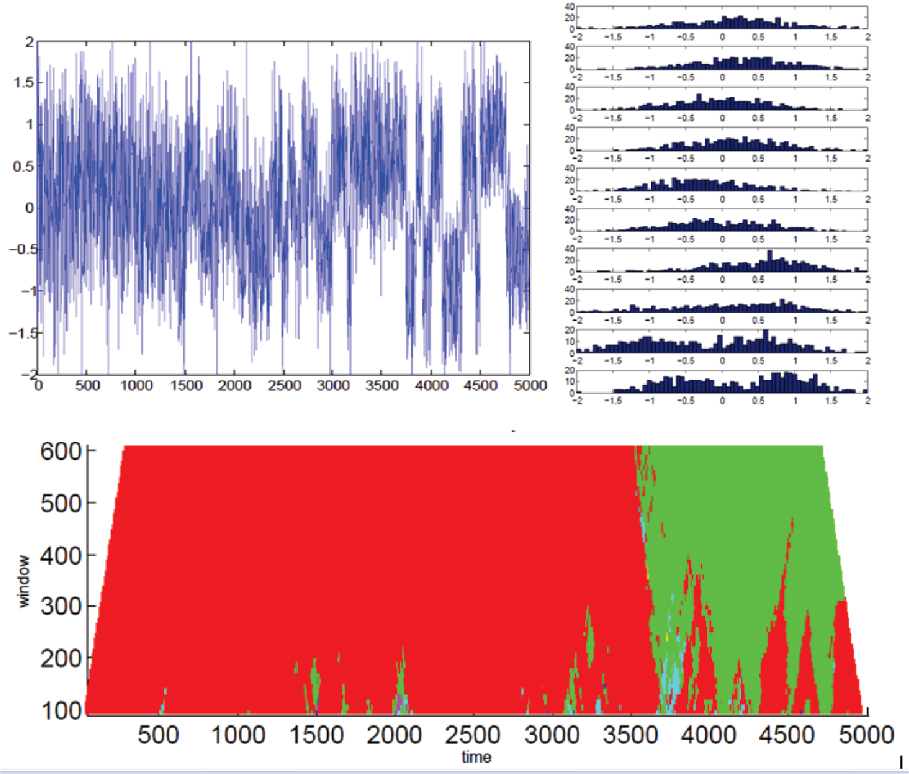}}
\caption{Test of potential analysis on artificial data from a system bifurcating from one state to two.
Here the underlying potential changes smoothly from one-well to double-well, described by the stochastic
potential equation with varying potential wells (10 chunks of 500 points each), with the bifurcation occurring
at time 3500 (a) artificial data generated from the changing potential function with a noise level 1;
(b) histograms of 10 chunks of data, from top to bottom, corresponding to consequent subsets of the series
(c) Contour plot of number of detected states, where red = 1 detected state, green = 2. Results plotted as a
function of sliding window length at the middle of the window.
}
\label{s2}
\end{figure}

We detect the order of the polynomial and hence the number of system states following the
method in~\citep{b10,b11}, plotting the results as a function of window length at the end of
each sliding window in a colour contour plot (e.g. Figure~\ref{f2}b).
The rate of correct detection
depends on sliding window size~\citep{b11}: when the window contains more than 400 data points
(which in the case of daily sea-ice data corresponds to about 1.1 years),
the success rate is 80\%, even when noise level is up to five times bigger than the depth of
the potential well; for larger windows it approaches 98\%.
A test of the method on artificial data, generated from a model system
in which the underlying potential bifurcates from one state to two,
illustrates correct detection of the number of system attractors (Figure~\ref{s2}).

Such tests employ Gaussian white noise, whereas sea-ice data is correlated.
In the case of correlated data, the probability density under investigation is the same (a pdf aggregates
data without taking into account its temporal organization).
However, correlated data is more likely to sample another state due to drift
(red noise) than is white noise.
Hence correlated data may have better detection rate statistics than uncorrelated data.

We derive the coefficients describing the shape of the potential using an unscented
Kalman filter~\citep{b10,b11}, while we estimate the noise level using wavelet de-noising with
Daubechies wavelets of 4th order~\citep{b11}.

The method assumes each subset of data is quasi-stationary and the noise is Gaussian white.
For the 4-year intervals used to reconstruct the potentials in e.g. Figure~\ref{f2}c, the assumption of
stationarity is reasonable. The noise in geophysical systems may be red rather
than white, but the assumption of white noise can still be valid provided that the noise
is stationary (DFA fluctuation exponent less than 1). By applying the potential model
in such cases, we may attribute part of the noise variability to the potential dynamics
when analysing the two components of the potential model. This model is an approximation;
still it allows us to derive accurately the structure of the potential for systems with
stationary red noise. When there are no non-stationarities such noise cannot artificially
create an additional system state.

\subsection{Critical slowing down}
To test for bifurcation in the residual sea-ice fluctuations we look for the signal
of `critical slowing down' beforehand~\citep{b15}.
Namely, for a low-order dynamical system approaching a bifurcation where its current state becomes unstable, and it
transitions to some other state, one can expect to see it become more sluggish in
its response to small perturbations~\citep{b15}. This can hold even for complex systems such
as the sea-ice, if they exhibit a bifurcation point, because near to it their behavior will reduce down to
that of a low-order system (following the Center Manifold Theorem). The signal of
`critical slowing down' is detectable as increasing autocorrelations in time series
data, occurring over timescales capturing the decay of the major mode in the system~\citep{b12},
which is controlled by the leading eigenvalue.  We looked for this early warning
indicator in the form of rising lag-1 autocorrelation~\citep{b12} (ACF-indicator), and through
detrended fluctuation analysis (DFA-indicator) as a rising scaling exponent~\citep{b13}.
Parabolic trends were removed prior to estimating these two indicators
(previously termed `propagators')
of critical slowing down. This is because any trend
affects autocorrelations and hence may cause false positive signals in the indicators.
To test robustness we also performed an alternative pre-processing of data; first removing
the quadratic downward trend and then deseasonalising the data, and obtained equivalent results.

\subsubsection{ACF-indicator}
Lag-1 autocorrelation was estimated~\citep{b12,b13}
by fitting an autoregressive model
of order 1 (linear AR(1)-process) of the form:
\begin{equation}
z_{t+1} = c \cdot z_t + \sigma\eta_t,									
\label{eq4}
\end{equation}
where $\eta_t$ is a Gaussian white noise process of unit variance,
and the `ACF-indicator' (AR1 coefficient):
\begin{equation}
c = e^{-\kappa\Delta t},											
\label{eq5}
\end{equation}
where $\kappa$ is the decay rate of perturbations, and $\kappa\to 0$ (i.e. $c\to 1$)
as bifurcation is approached~\citep{b12}.

\subsubsection{DFA-indicator}
Detrended fluctuation analysis (DFA) extracts the fluctuation function of
window size $s$, which increases as a power law if the data series
is long-term power-law correlated:
\begin{equation}
F(s)\propto s^\alpha											
\label{eq6}
\end{equation}
where $\alpha$ is the DFA scaling exponent.  In the short-term regime,
as $c\to 1$ of the AR(1)-model,
the slowing
exponential decay is well approximated by a power law in which $\alpha\to 1.5$,
in the time interval 10-100 units.  Exponent $\alpha$
is rescaled, following~\citep{b13}, to give a `DFA-indicator'
that reaches $1$ at critical behavior.

\subsubsection{Variance}
We also monitored variance (calculated as standard deviation),
because if a state is becoming less stable this can be characterized by its
potential well getting shallower, causing increased variability over time
(although this is not independent of lag-1 autocorrelation~\citep{b26}).

\subsubsection{Indicator trends}
Upward trends in the indicators (rather than their absolute value)
provide the primary early warning signal.
The Kendall $\tau$ rank correlation coefficient~\citep{b30} measures
the strength of the tendency of an
indicator to increase (positive values) or decrease
(negative values) with time, against
the null hypothesis of a random sequence of measurements
against time (value approximately zero).
As a sensitivity analysis, the sliding window along the
time series was varied from 1/4 to 3/4 of
the series length.

\subsection{Model of increasing seasonal cycle}
To examine whether the results obtained could be explained by an
increase in the amplitude of the seasonal cycle~\citep{ditlev},
we built a simple stochastic model, described by the following equation:
$$
\begin{array}{c}
x(t)=L+A\cdot\sin\left(\frac{2\pi t}{365}  \right)    + \sigma\eta,\\[10pt]
A=\left\{
\begin{array}{l}
1,\ \  \mbox{when}\ t=1, \, \dots \, , 7300,\\[10pt]
\frac{0.5}{12410-7300}t+\frac{12410-1.5\cdot 7300}{12410-7300},\\[10pt]
\mbox{when}\ t=7301,\, \dots\, , 12410,
sign\left(\sin\left(\frac{2\pi t}{365}  \right)\right)<0,
\end{array}
\right.
\end{array}
$$
where $\eta$ is Gaussian white noise of unit variance, $\sigma=0.15$.
This simulates sinusoidally varying `daily' data (period 365)
over a period $t=1:12410$ corresponding to 34 years, equivalent
to period 1979-2012.
A global declining trend
of the data is simulated as linear in form; $L=-0.02\cdot t + 55.82$.
Also the amplitude of the lower half of the sine wave starts to grow
linearly after 20 `years' of simulated data (i.e. in `year' 1998),
such that it changes from -1 to -1.5 at the end of time series.

The model data was pre-processed similarly to the sea-ice data;
We first deseasonalised the model data (removing 365-day `seasonal' average)
and performed potential analysis of the residuals.
We then removed a quadratic trend from the series
and calculated the early warning indicators.

\section{Results and Discussion}

\subsection{Multi-modality detection}
\begin{figure}[h!]
\centerline{\includegraphics[width=.6\textwidth]{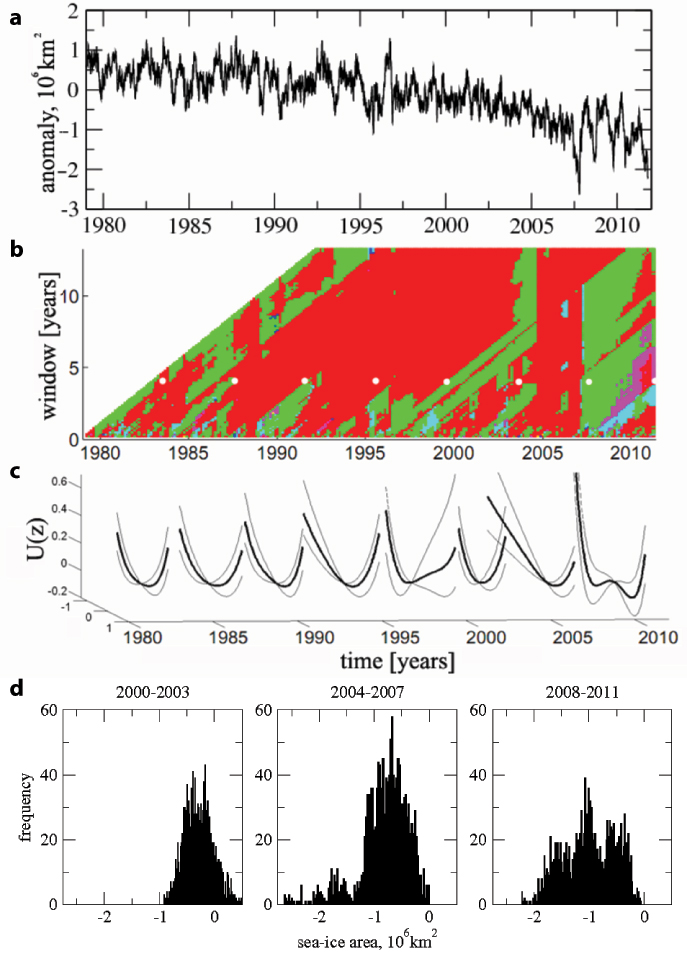}}
\caption{Analysis of Arctic sea-ice area. (a) Sea-ice area anomaly,
daily data with mean seasonal cycle removed. (b) Contour plot of number of detected states,
where red = 1 detected state, green = 2, cyan = 3, magenta = 4. Results plotted as a
function of sliding window length at the end of the window. (c) Reconstructed potential
curves of eight 4-year time intervals, corresponding to the white dots in (b). Here 'z'
is sea-ice area fluctuation on a shifted scale. Faint lines are potential curves derived
from error estimates on the coefficients of the polynomial potential function
(for details see ref. 11).
In the penultimate interval 2004-2007 a second state starts to appear
and in the final interval 2008-2011 there are two states of comparable stability.
(d) Histograms of the data for 2000-2003, 2004-2007, 2008-2011
from which the corresponding potential curves are derived (see Methods).}
\label{f2}
\end{figure}

After removing the mean seasonal cycle (1979-2008), the
remaining fluctuations in sea-ice area
include some of order 10$^{6}$ km$^{2}$ (Figure~\ref{f2}a).
The largest anomalies are in 1996 (maximum of the series) and 2007-2011 (minima).
They typically occur in the summer-autumn, when the sea-ice area is at
its lowest in the seasonal cycle.
Given the size of sea-ice fluctuations during 2007-2011 (Figure~\ref{f2}a)
and the pronounced drop in sea-ice minima relative to
sea-ice maxima since 2007 (Figure~\ref{f1}c),
we considered whether
the residuals exhibited an abrupt change to multi-modality in 2007.

On analyzing the residual sea-ice area fluctuations using our method of potential analysis,
over long time windows (here $>$1 year),
we typically find a single mode and corresponding attractor, representing the
normal seasonal cycle of sea-ice variability (Figure~\ref{f2}b). Sometimes a second mode is
detected associated with e.g., the sea-ice maximum in 1996, but these changes
are not found simultaneously and persistently across a wide range of window lengths.
However, from 2007 onwards, a persistent switch to two modes or attractors
is detected, across a wide range of window lengths up to $>$10 years (Figure~\ref{f2}b).
The same switch
is also detected in analysis of the shorter record of sea-ice extent data (Figure~\ref{s3}b).

The stability of the attractor(s) for the residual sea-ice fluctuations can be reconstructed, in the form of
potential curves for fixed intervals of the data (Figure~\ref{f2}c), with associated
error estimates (on the coefficients of the polynomial function describing the
potential~\citep{b11}). The sea-ice residuals are typically characterized by a single mode and corresponding attractor.
The interval 1996-9 (including the 1996 maximum anomaly) shows signs of a
second higher ice cover attractor that is degenerate (i.e. not fully stable). In 2000-3
there is a return to a single attractor.
In 2004-7, which includes
the record September 2007 sea-ice retreat, a low ice-cover attractor starts to appear
in the fluctuations.
Then in 2008-11 the potential separates into two attractors, although the error
range allows for one or the other of these to be degenerate.

The potential curves
are derived from histograms of the original data~\citep{b11} (Figure~\ref{f2}d), which confirm a
second mode appearing among a long tail of negative fluctuations during 2004-2007,
followed by a separation of multiple modes during 2008-2011, which the method fits as a bi-modal distribution.
Thus, we originally hypothesized that the Arctic sea-ice recently passed a bifurcation point~\citep{cryo},
which created a new lower ice cover attractor for the residual deseasonalised fluctuations.
Since then it has fluctuated between its normal attractor for seasonal
variability and the new, lower ice cover attractor.

\begin{figure}[h!]
\centerline{\includegraphics[width=0.9\linewidth]{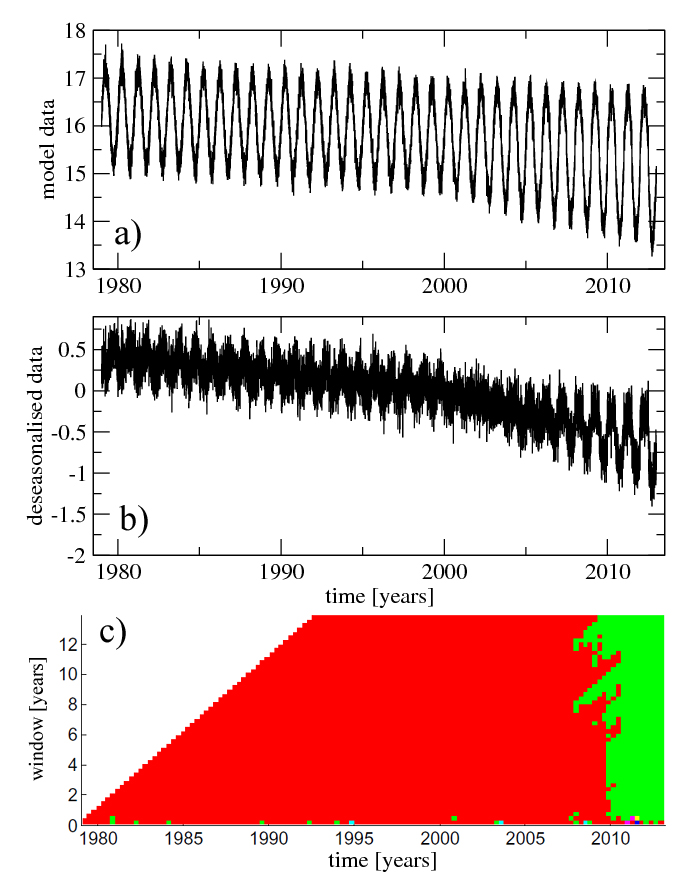}}
\caption{Model `daily' data with an overall decline and increase
in the amplitude of the lower half of the
sine wave `seasonal cycle' as described in the text:
(a)~raw data, (b)~deseasonalised data and (c)~contour plot of the number
of detected states in the deseasonalised data.}
\label{dit1}
\end{figure}

However, an abrupt change in the amplitude of the seasonal cycle will leave a residual record
that has some seasonality on one side of the transition or the other~\citep{ditlev}.
These remnant seasonal fluctuations will in turn produce a bi-modal distribution,
which is accurately detected by our method -- hence care is needed over how to interpret this.
Sure enough analysis of the stochastic model of an increasing seasonal cycle,
shows some qualitatively similar results (Fig.~\ref{dit1}) to analysis of the real sea-ice data.
In the model, imposed growth in the amplitude in the lower half of the seasonal cycle has been underway for
over a decade before the residuals are detected as bi-modal.
In contrast, in the real data the increase in amplitude of the seasonal cycle is abrupt (Fig.~\ref{f1}c)
and is detected immediately (Fig.~\ref{f2}).

\subsection{Early warnings?}
Having hypothesized that a bifurcation may have occurred in Arctic sea-ice cover,
we tested this by examining whether it was preceded (or followed)
by any signals of destabilization in the form of critical slowing down.
However, a caveat here is that the inferred bifurcation
(Figures~\ref{f2}, \ref{s3}), if correct, represents the
creation of a new ice cover attractor (for the residual fluctuations)
rather than the total loss of stability of the existing ice cover attractor.
Hence the existing ice cover attractor may not show clear destabilization
prior to the bifurcation.

\begin{figure}[h!]
\centerline{\includegraphics[width=.9\textwidth]{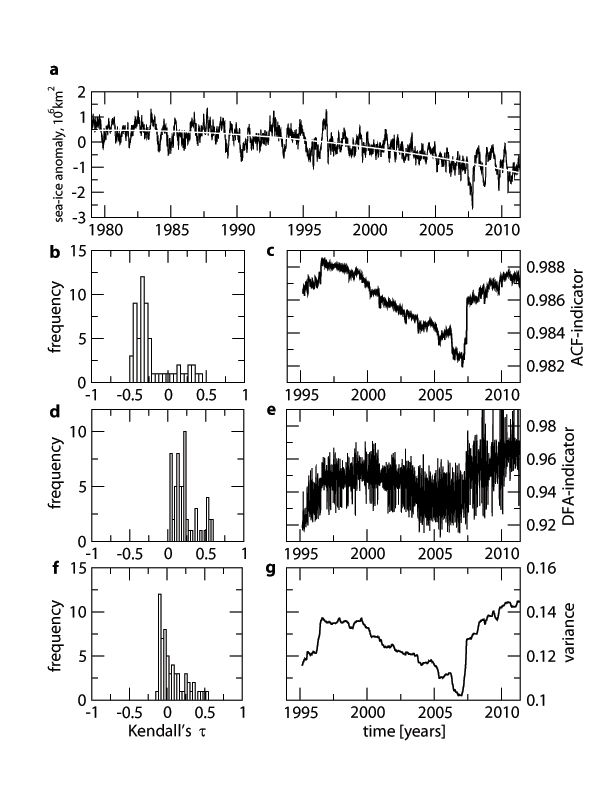}}
\caption{Search for early warning signals of bifurcation in Arctic sea-ice area data.
(a)~Sea-ice area anomaly (as in Figure 2a) showing the quadratic downward trend that
is removed prior to calculating the instability indicators. Right panels show example
indicators using a sliding window of length half the series, with results plotted at
the end of the sliding window. Indicators from: (c)~autocorrelation function (ACF),
(e)~detrended fluctuation analysis (DFA) and (g)~variance. Left panels show histograms
of the Kendall   statistic for the trend in the indicators when varying the sliding
window length from 1/4 to 3/4 of the series: (b)~ACF-indicator,
(d)~DFA-indicator, (f)~variance.}
\label{f3}
\end{figure}

Prior to 2007 there is no consistent early
warning signal of destabilization (Figure~\ref{f3}c,e,g). The indicators all increased
around the anomalous sea-ice maximum in 1996, but then they all declined toward 2007,
consistent with our potential reconstruction (Figure~\ref{f2}c).
The only early warning signal prior to 2007 is a rise in the DFA-indicator
in analysis of sea ice extent (Figure~\ref{s4}). Sensitivity analysis confirms
this is the only robust increase across the three indicators and the two datasets,
prior to 2007 (Figure~\ref{s5}). Thus, there was no consistent early warning
signal of critical slowing down before the hypothesized bifurcation.
Instead the sea-ice showed signs of increasing stability in the preceding decade,
contrary to what would be expected from an approach to bifurcation.

The sea-ice retreat in 2007 caused abrupt increases in all the indicators, which
have continued to rise since then (Figure~\ref{f3}c,e,g). Sensitivity analysis reveals
a robust upward trend in the DFA-indicator across the whole dataset (Figure~\ref{f3}d),
but no robust overall trend in the ACF-indicator or variance
(Figure~\ref{f3}b,f). These results are reproduced
in analysis of the shorter record of sea-ice extent data (Figure~\ref{s4}).
The rise in the DFA-indicator could be consistent with the sea-ice
having increasing `memory' of its earlier states due to
critical slowing down~\citep{b13}. The somewhat
different behavior of the ACF and DFA indicators
could then be explained by the different time scales used for their calculation. The ACF-indicator,
based only on lag-1 autocorrelation (here from one day to the next), may be monitoring the
behavior of fast decay modes unrelated to critical slowing down.  The DFA-indicator in
contrast is calculated on time scales up to 100 days, which should be long enough to
capture the slowest recovery mode of the sea-ice.

We conclude that overall the indicators detect a profound shift in the data
in 2007, but do not forewarn of it. This does not convincingly support the bifurcation interpretation.
Since 2007 an ongoing destabilization is detected.

\begin{figure}[h!]
\centerline{\includegraphics[width=0.9\linewidth]{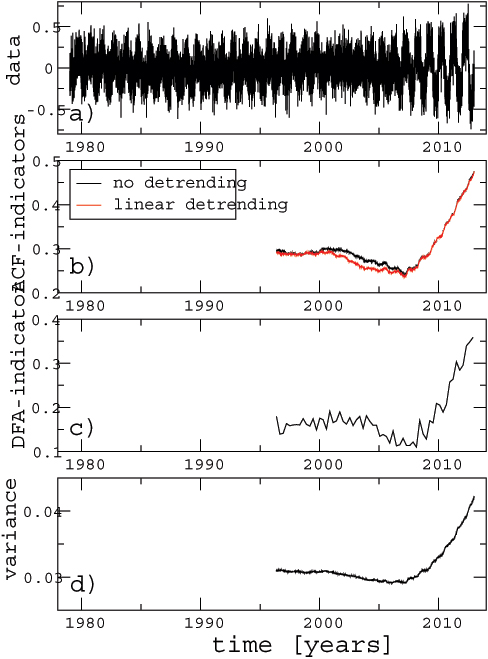}}
\caption{(a)~Detrended deseasonalised model data (Figure~\ref{dit1})
and its early warning indicators: (b)~ACF-indicators (with and without linear detrending
within sliding windows); (c)~DFA-indicator; (d)~variance.}
\label{dit2}
\end{figure}

The stochastic model of an increasing seasonal cycle shows no clear trend in
the ACF or DFA indicators or the variance, followed by a steady rise
in all the indicators as the residual data becomes bi-modal (Fig.~\ref{dit2}).
However, there are no abrupt increases in the ACF-indicator of critical slowing down or the variance (Fig.~\ref{dit2}),
as there are in analysis of the sea-ice data around 2007 (Fig.~\ref{f3}).
This is consistent with the change in amplitude of the seasonal cycle
being much more abrupt in the real data than in the model.

\subsection{Seasonal analysis}
Our results may be sensitive to the fact that land masses mute variations in winter-spring ice area~\citep{b24},
whereas summer-autumn area is less affected. To address this we analyzed a derived index
of `equivalent sea-ice extent'~\citep{b24}, which is based on the latitude of the sea-ice edge
where it is free to migrate, converted to an area, assuming there were no continents present.
Fluctuations are much larger in this index, and recent summer-autumn ice retreats no
longer stand out as anomalous~\citep{b24},
hence no recent recent shift from uni-modal to multi-modal residuals is detected
(Figure~\ref{s7}).
However, there is still a signal of overall destabilization (Figure~\ref{s8}),
which appears before the signal in actual sea-ice area (Figure~\ref{f3}).
This suggests the abrupt change in the data detected in 2007
could be (at least partly) a geographic property of the shrinkage of summer-autumn ice cover
away from the continents facilitating larger fluctuations~\citep{b24}.

\begin{figure}[h!]
\centerline{\includegraphics[width=.6\textwidth]{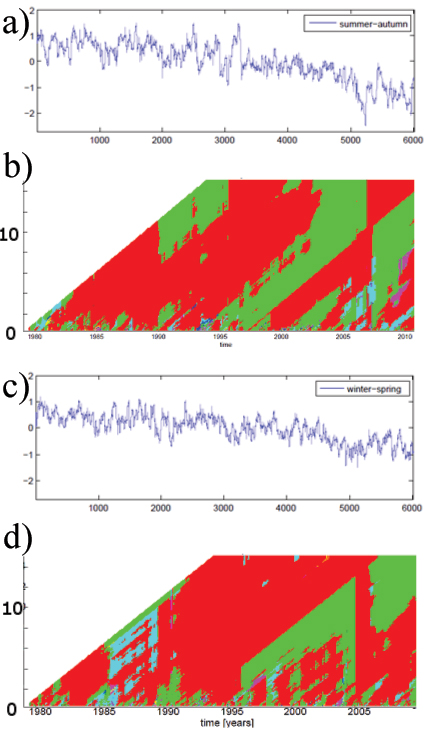}}
\caption{Potential analysis of summer-autumn and winter-spring Arctic sea-ice area data.
(a)~Summer-autumn sea-ice area anomaly, daily data with mean cycle removed. (b)~Contour plot
of number of detected states. (c)~Winter-spring sea-ice area anomaly, daily data with mean cycle
removed. (d)~Contour plot of number of detected states.
}
\label{s9}
\end{figure}

\begin{figure}[h!]
\centerline{\includegraphics[width=.9\textwidth]{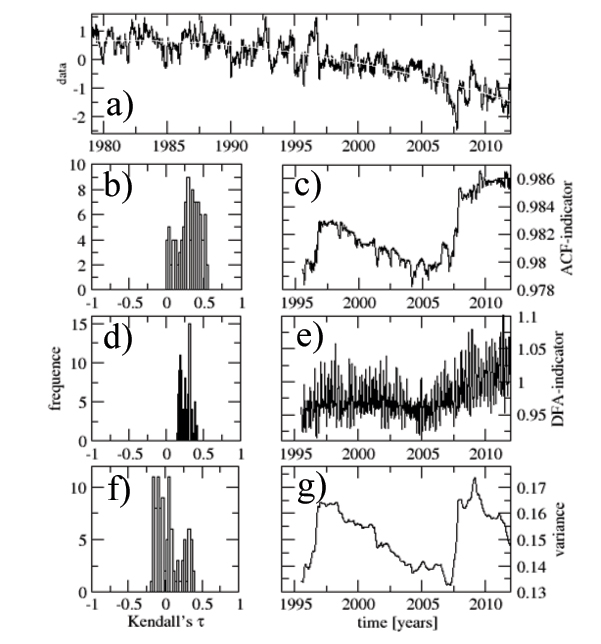}}
\caption{Search for early warning signals of bifurcation in
summer-autumn Arctic sea-ice area.
(a)~Summer-autumn sea-ice area anomaly
(as in Figure~\ref{s4}a) showing the quadratic downward trend that
is removed prior to calculating the instability indicators.
Right panels show example indicators from:
(c)~autocorrelation function (ACF), (e)~detrended fluctuation analysis
(DFA) and (g)~variance, results
plotted at end of a sliding window of length half the series.
Left panels show histograms of the Kendall
statistic for the trend in the indicators when varying the
sliding window length from 1/4 to 3/4 of the series:
(b)~ACF-indicator, (d)~DFA-indicator, (f)~variance.
}
\label{s10}
\end{figure}

\begin{figure}[h!]
\centerline{\includegraphics[width=.9\textwidth]{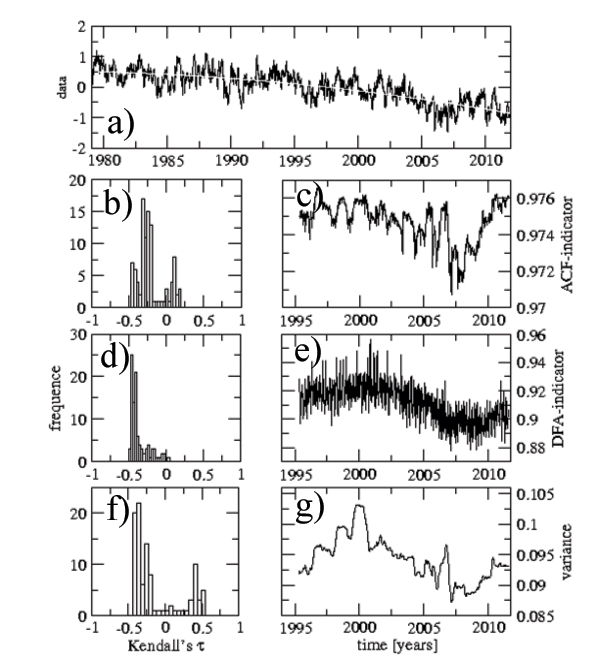}}
\caption{Search for early warning signals of bifurcation in
winter-spring Arctic sea-ice area.
(a)~Winter-spring sea-ice area anomaly (as in Figure~\ref{s4}c)
showing the quadratic downward trend that
is removed prior to calculating the instability indicators.
Right panels show example indicators from:
(c)~autocorrelation function (ACF), (e)~detrended fluctuation
analysis (DFA) and (g) variance,
results plotted at end of a sliding window of length half the series.
Left panels show histograms of
the Kendall statistic for the trend in the indicators when
varying the sliding window length from
1/4 to 3/4 of the series: (b)~ACF-indicator, (d)~DFA-indicator, (f)~variance.
}
\label{s11}
\end{figure}

To examine whether this is the case, we
subdivided the original data
(Figure~\ref{f1}a) into two composite series;
summer-autumn (June-November inclusive) and winter-spring (December-May inclusive),
removing the mean cycle from each, and re-running the analysis.
Both subsets of the data carry part of the signal of
abrupt change in 2007 (Figure~\ref{s9}),
suggesting the
change in the dynamics is not a purely summer-autumn phenomenon.
The signal is clearest in summer-autumn, but does not span as wide a range of window lengths
as in the full data analysis.
 However, the summer-autumn data does show upward trends
in the ACF and DFA indicators and (less clearly) the variance (Figure~\ref{s10}),
which are generally stronger than in the full dataset (Figure~\ref{f3}).
In contrast, the winter-spring data shows no convincing upward trends
in any of the indicators (Figure~\ref{s11}).
Thus, the recent signal of increasing auto-correlation and variance
(i.e. destabilization)
is associated primarily with summer-autumn
sea-ice fluctuations.

\subsection{Summary and geophysical mechanisms}
An abrupt and persistent change in sea-ice dynamics is detected to have occurred in 2007.
This involves an extra $\sim$10$^{6}$ km$^{2}$ or more sea-ice loss each summer-autumn since then.
Our initial hypothesis that this abrupt increase in the amplitude of the seasonal cycle of sea-ice variability
occurred through a bifurcation mechanism~\citep{cryo}, is not consistently supported.
Thus, the underlying causal mechanism remains uncertain.
Still, there must be some amplifying positive feedback mechanisms contributing to the abrupt increase in
summer-autumn ice loss.

Statistical models such as ours cannot shed light on these underlying geophysical mechanisms.
However,
several positive feedbacks have been identified in recent data and are worth mentioning.
Sea-ice retreat since 1979
has exposed a dark ocean surface, causing 85\% of the Arctic region to receive an increase in
solar heat input at the surface, with an increase of 5\% per year in some regions~\citep{b16}.
This is warming the upper Arctic ocean and contributing to melting on the bottom of the
sea-ice~\citep{b17}.  Sea-ice retreat is also amplifying warming of the lower atmosphere in the
Arctic~\citep{b18}, which is shifting precipitation from snow to rainfall,
and where rain lands on the remaining
sea ice cover, it is encouraging melt~\citep{b19}.
The loss of multi-year ice thins the average ice cover making it more vulnerable to
further summer losses~\citep{comiso}.
Finally, sea-ice loss is beginning to change
atmospheric circulation patterns~\citep{b28} (although how
that feeds back to ice cover is unclear).

The abrupt increase in the seasonal cycle that
we detect clearly does not involve total seasonal
sea-ice loss and hence is sub-Arctic in scale.
However, there may be a precedent for this;
past abrupt Arctic cooling and warming events
have been linked to switches between alternative states
for sea-ice cover in the Barents and
Kara Seas region~\citep{b20,b21}. Such sub-Arctic-scale
switches can still have significant impacts,
indeed recent ice loss from the Barents and Kara Seas has
been linked to cold winter extremes
over Eurasia~\citep{b29}.
The connection between surface
temperature, sea level pressure and winds in the Arctic
region, and their effect on the sea-ice cover is discussed
by~\citep{comiso}.

\medskip

\noindent
{\bf Conclusions.}
We detect an abrupt and persistent increase in the amplitude of seasonal sea-ice variation in 2007.
This involves an extra $\sim$10$^{6}$ km$^{2}$ or more sea-ice loss each summer-autumn then and since.
We originally hypothesized that this abrupt change could be explained in terms of a bifurcation
in which a new lower ice cover attractor (for deseasonalised sea-ice fluctuations)
appeared and began to be sampled in every summer-autumn from 2007 onwards.
However, this interpretation is clearly sensitive to how the seasonal cycle is removed from the raw data,
and also to the presence of continental land masses restricting winter-spring ice fluctuations.
Furthermore, there was no robust early warning signal of critical slowing down,
as would be expected prior to the hypothesized bifurcation.
Early warning indicators do however show destabilization of the summer-autumn
sea-ice cover since 2007.
Overall, the bifurcation hypothesis lacks consistent support.
Instead we can say that there has been an abrupt and persistent jump
in the amplitude of the seasonal cycle of Arctic sea-ice cover in 2007~\citep{ditlev},
but the underlying causal mechanism remains uncertain.
We describe this as a (non-bifurcation) `tipping point',
because it involved an abrupt, qualitative change in the sea-ice dynamics,
without any evidence for a large forcing perturbation, i.e. the abruptness
resides in the internal dynamics of the Arctic climate system.

Our statistical methods detected this `tipping point' and its time of onset
suggesting they might usefully be applied to real-time
analysis of diverse climatological data
(albeit in this case the change retrospectively appears fairly clear in the raw data).
However, statistical methods cannot shed light on geophysical mechanisms.
To make progress on the underlying causal mechanisms requires process-based models.
Potentially the statistical indicators of stability could be used to help
re-calibrate the sensitivity of process-based models, which have
generally proved unable to capture the observed abruptness of decline of the Arctic sea-ice cover.

\medskip

\noindent
{\bf Acknowledgements}
The research was supported by the NERC project
'Detecting and classifying bifurcations in the climate system'
(NE/F005474/1) and by the AXA Research Fund through a postdoctoral
fellowship for V.N.L.
Research was carried out on the High Performance Computing
Cluster at the University of East Anglia.
We thank J. Imbers Quintana and A. Lopes for discussions at the outset of this work,
M. Scheffer and P. Ditlevsen for discussions over the interpretation of the results,
and two anonymous reviewers for their robust critiques of the Discussion paper.
The research was carried out on the High
Performance Computing Cluster supported by the Research Computing Service at
the University of East Anglia.




\begin{figure}[h!]
\centerline{\includegraphics[width=.9\textwidth]{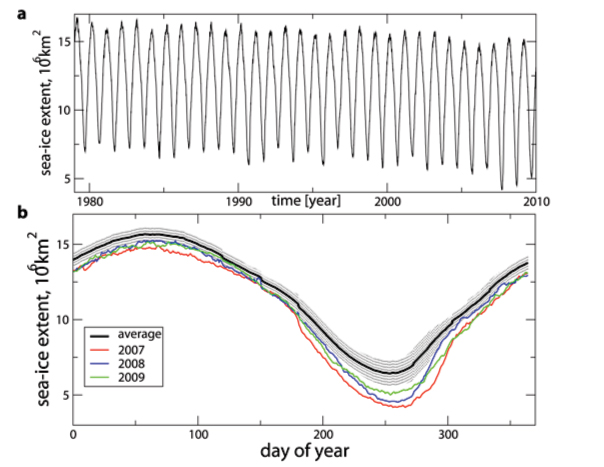}}
\caption{Arctic sea-ice extent from satellite data.
(a) Arctic sea-ice extent from Eisenman~\citep{b24}
ftp://ftp.agu.org/apend/gl/2010gl043741. (b) The mean
annual cycle of the extent data over 1979-2009
(solid line, shaded area denotes $2\sigma$ error bars),
together with the last three anomalous years.}
\label{s1}
\end{figure}

\begin{figure}[h!]
\centerline{\includegraphics[width=.6\textwidth]{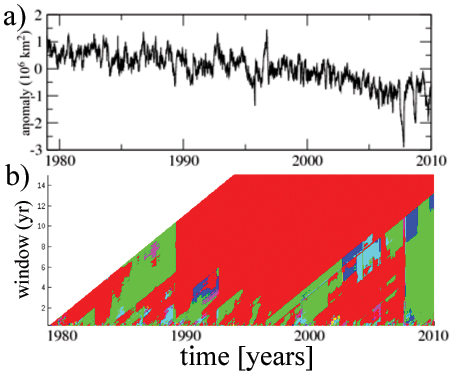}}
\caption{Analysis of Arctic sea-ice extent. (a) Sea-ice extent anomaly, daily
data with mean seasonal cycle removed. 
(b) Contour plot of number of detected states, where red = 1 detected
state, green = 2, cyan = 3, magenta = 4. 
Results plotted as a function of sliding window length at the end
of the window.}
\label{s3}
\end{figure}
\begin{figure}[h!]
\centerline{\includegraphics[width=.9\textwidth]{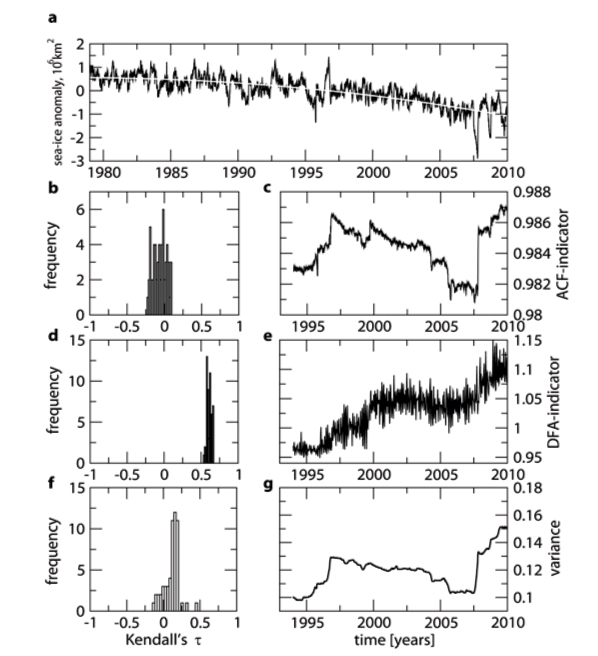}}
\caption{Search for early warning signals of bifurcation in Arctic sea-ice extent data.
(a) Sea-ice extent anomaly (as in Figure~\ref{s2}a) showing the quadratic downward trend that is
removed prior to calculating the instability indicators. Right panels show example indicators from:
(c) autocorrelation function (ACF), (e) detrended fluctuation analysis (DFA) and (g) variance,
results plotted at end of a sliding window of length half the series. Left panels show histograms of
the Kendall statistic for the trend in the indicators when varying the sliding window length from
1/4 to 3/4 of the series: (b) ACF-indicator, (d) DFA-indicator, (f) variance.
}
\label{s4}
\end{figure}
\begin{figure}[h!]
\centerline{\includegraphics[width=.9\textwidth]{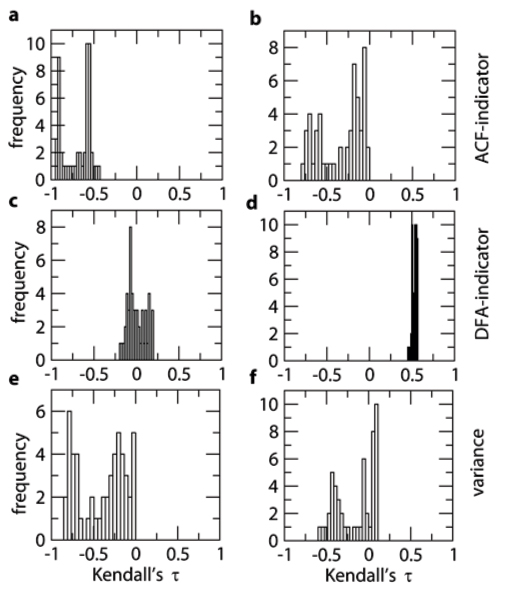}}
\caption{Destabilisation indicators calculated up to 2007. From; (a,c,e) sea-ice area anomaly,
(b,d,f) sea-ice extent anomaly (both after detrending). Sensitivity analysis when varying sliding
window length for Kendall trend statistic of: (a,b) ACF-indicator (c,d) DFA-indicator (e,f) variance.
}
\label{s5}
\end{figure}

\begin{figure}[h!]
\centerline{\includegraphics[width=.9\textwidth]{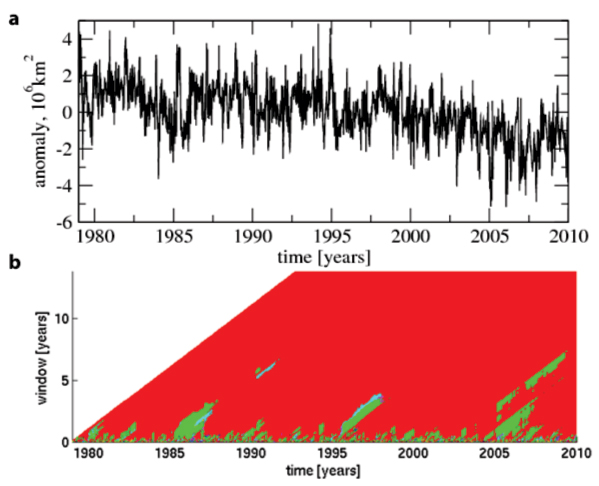}}
\caption{Potential analysis of equivalent sea-ice extent index.
(a)~Dataset constructed by Eisenman~\citep{b24}
and available at: ftp://ftp.agu.org/apend/gl/2010gl043741,
which is based on the latitude of the
Arctic sea ice edge where the ice is free to migrate,
converted to an equivalent area, assuming there
were no land masses in the high northern latitudes.
(b)~Contour plot of number of detected states,
where red = 1 detected state, green = 2, cyan = 3,
magenta = 4. Results plotted as a function of
sliding window length at the end of the window.
No bifurcation is detected in this dataset, because
it has much higher internal variability
than sea ice extent (Figure~\ref{s4}(g)) and recent
observed ice extent anomalies are dwarfed by earlier,
larger fluctuations that are inferred would
have occurred had the continents not got in the way of winter ice variations.
}
\label{s7}
\end{figure}
\begin{figure}[h!]
\centerline{\includegraphics[width=.9\textwidth]{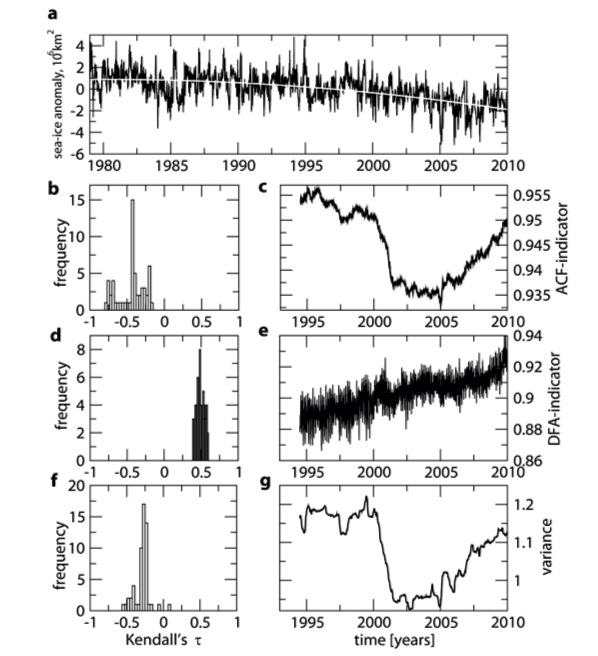}}
\caption{Search for signals of destabilisation in equivalent
sea-ice extent. (a)~Equivalent sea-ice
extent index (as in Figure~\ref{s3}) showing the quadratic
downward trend that is removed prior to
calculating the instability indicators. Right panels show
example indicators from: (c)~autocorrelation
function (ACF), (e)~detrended fluctuation analysis (DFA)
and (g)~variance; results plotted at end of
a sliding window of length half the series. Left panels
show histograms of the Kendall statistic for
the trend in the indicators when varying the sliding
window length from 1/4 to 3/4 of the series:
(b)~ACF-indicator, (d)~DFA-indicator, (f)~variance.
}
\label{s8}
\end{figure}

\end{document}